\documentclass[12pt,preprint]{aastex}

\slugcomment{}

\shorttitle{More evidences for CSL-1} \shortauthors{Sazhin et al.}

\begin{document}

\title{Further spectroscopic observations of the CSL-1 object}

\author{M. Sazhin\altaffilmark{1}, M. Capaccioli\altaffilmark{2,3}, G. 
Longo\altaffilmark{3,2,4}}

\and

\author{M. Paolillo\altaffilmark{3,4}, O. Khovanskaya\altaffilmark{1}}

 \affil{1 - Sternberg Astronomical Institute, Moscow State University,
University pr. 13, Moscow, RUSSIA}
 \affil{2 - INAF - Osservatorio
Astronomico di Capodimonte,via Moiariello 16, 80131, Napoli ITALY}
 \affil{3 - Department of Physical Sciences, University of Napoli Federico 
II,via Cinthia 9, 80126, ITALY}
 \affil{4 - INFN, Napoli Unit, ITALY}

\email{sazhin@sai.msu.ru}

\begin{abstract}

\noindent CSL-1 is a peculiar object (R.A.$_{2000}= 12^h \ 23^m \ 30"5$; 
$\delta_{2000}= -12^{\circ}\ 38' \ 57"0$) which, for its photometric and 
spectroscopic properties, is possibly the first case of gravitational lensing 
by a cosmic string. 
In this paper we present additional evidences, based on medium-high resolution 
VLT + FORS1 observations, that the spectra of the two components of CSL-1 are 
identical within a confidence level higher than 98$\%$ and the velocity difference of the 
two components is consistent with zero. 
This result adds further confidence to the interpretation of the system as a 
true lens.
\end{abstract}


\section{Introduction}
As recently stated in the masterly review by \citet{kibble}, the last two 
years have seen a renewal of interest in the cosmological role of cosmic 
strings, after more than a decade of relative quiescence: 
an interest which was mainly triggered by the discovery of
the unusual object CSL-1 (R.A.$_{2000}= 12^h \ 23^m \ 30"5$;
$\delta_{2000}= -12^{\circ}\ 38' \ 57"0$) \cite[][hereafter Paper I]{csl1} 
in the OAC-DF
(Osservatorio Astronomico di Capodimonte - Deep Field; \citealp{cap2})
and by the indirect evidence
obtained by \cite{schild} from the luminosity fluctuations of
the quasar $Q0957+561 \ A,B$.

\begin{figure}
\includegraphics*[width=7cm]{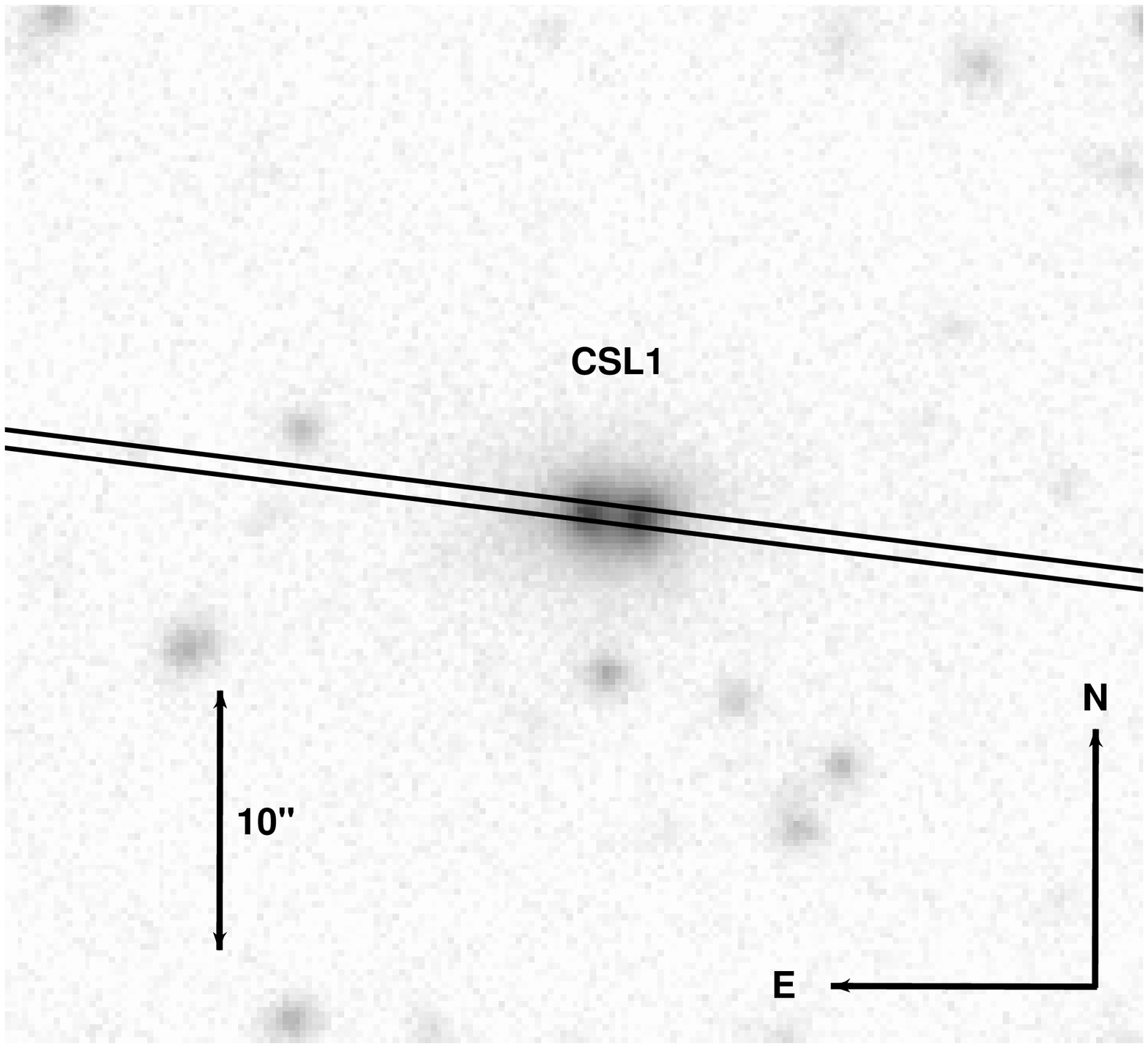}
 \caption{The CSL-1 field with the slit position drawn upon it.}\label{CSL1}
\end{figure}

CSL-1 (see Fig.\ref{CSL1}) is a double source laying in a
low density field.
In the original images the components, 1.9 arcsec apart, appeared to be 
extended and with roundish and identical shapes. 
By low resolution spectroscopy we learned that both components are at a redshift 
of $0.46\pm 0.008$, and by photometry (both global properties and luminosity 
profiles) that they match the properties of two giant elliptical galaxies. 
Detailed analysis showed that the spectra of the two components were
identical at a 99.9\% level. 
Such a conclusion, however, was hindered both by the limited wavelength range 
spanned by the spectra, and by their relatively low signal-to-noise ratio.

As discussed in Paper I, the only possible explanation of the CSL-1
properties is: {\it i)} either an unlikely chance alignment of two
giant ellipticals at the same redshift and with very similar spectra, or
{\it ii)} a gravitational lensing phenomenon. 
But in this second case, due to the lack of asymmetry in the two images, the 
lens could not be modelled with the standard lensing by a massive compact 
source. 
Actually, the usual gravitational lenses, i.e. those formed by bound clumps of 
matter, always produce inhomogeneous gravitational fields which distort the 
images of extended background sources \cite[cf.][]{gl,kee}. 
The detailed modelling of CSL-1 proved that the two images were virtually 
undistorted (see Paper I for details).
The only other explanation left in the framework of the gravitational lensing
theory was that of a lensing by a cosmic string.

In Paper I we indicated two possible {\it experimenta crucis}: {\it i)} the 
detection of the sharp edges at faint light levels, since this is the signature 
expected from the lensing by a cosmic string, and {\it ii)} the detection of a 
small amplitude but sharp discontinuity in the local CMB \citep{gangui}. 
The first test calls for high angular resolution and deep observations with HST; 
time has already been allocated but the observations have not yet been 
performed. 
The second test was attempted using the preliminary release of the WMAP data 
\citep{kai84, WMAP}. 
By careful processing, a $1 \sigma$ positive detection at the position of CSL-1 
was found. 
Even though such a detection seems to imply the unrealistic speed of $0.94 \ c$ 
for the string, the authors pointed out that both the low angular resolution and 
the low S/N ratio could prevent the detection of the expected signature.

Another test was proposed by several authors \citep{string1, string2, string3, string4}, 
and more recently in \citet{string5}.
It is based on the fact that the alignment of the background object (a galaxy) 
inside the deficit angle of the string is a stochastic process determined by the 
area of the lensing strip and by the surface density distribution of the
extragalactic objects which are laying behind the string. 
All the lensed objects will fall inside a narrow strip defined by the
deficit angle computed along the string pattern. 
A preliminary investigation of the CSL-1 field showed a significant excess of
gravitational lens candidates selected on the bases of photometric
criteria only \citep{saz04}. 
Spectroscopic observations are being obtained for a first set of candidates and 
will be discussed elsewhere.

\noindent In this paper we address on firmer grounds the issue of the 
gravitational lens nature of CSL-1 using intermediate resolution spectra obtained 
at the ESO Very Large Telescope + FORS 1 spectrograph.

\section{The data and data reduction}
New spectra of CSL-1 were obtained in March 2005 at VLT\footnote{The Very Large 
Telescope is operated by the European Southern Observatory and is located at 
Mount Paranal in Chile; http://www.eso.org/paranal/} using the FORS 1
spectrograph under Director's Discretionary Time (proposal 274.A-5039).

The spectra were acquired on March 15-19 2005
in the FORS 1 long slit configuration 600V+GG435 ($\lambda/\Delta\lambda=990$ at central 
wavelength), setting the slit of the spectrograph across the centers of the two components of 
CSL-1. The observations were split in several exposures, to prevent saturation and to 
allow for a better removal of the bad pixels and cosmic rays.
The risk of cross contamination between the spectra of the two components was 
minimized by retaining only the 6 frames with an average Signal to Noise (S/N) ratio of $\sim 
12$ and a PSF FWHM$<1.0"$\footnote{The Point Spread Function represents the image of a pointlike object. 
Its Full Width at Half Maximum measures the level of blurring due to atmospheric and instrumental factors.}, for a total exposure time of 4740 sec. 

\begin{figure}
\includegraphics*[width=8cm]{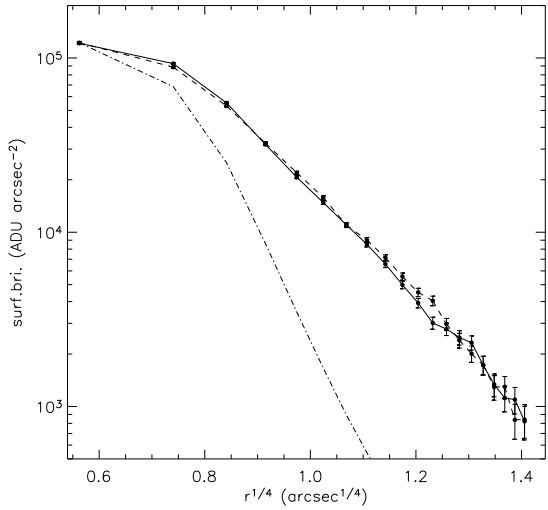}
 \caption{The integrated light profiles for the two components of CSL-1
(solid and dashed line) compared to the light profile of an unresolved source
(dot-dashed line). Galaxy profiles were extracted avoiding the region of overlap,
 i.e a sector of $60^\circ$ toward the other component. Errorbars represent $1\sigma$
 uncertainties.}
\label{lightprofile}
\end{figure}

During observations a short $R$-band exposure was also obtained to check the 
pointing of the instrument. 
Owing to the excellent seeing conditions, this image could be used to test the
extended nature and the similarity of the light profiles of the two images. 
The dot-dashed line (with no data symbols) in Figure~\ref{lightprofile} shows 
the light profile of an unresolved star present in the field, compared to the 
light profiles of the two images of CSL-1, which appear to be clearly resolved 
and identical within the errors. A de Vaucouler fit yields $r_e\simeq 1.6"$.

The spectral data were reduced through standard MIDAS procedures
and, after bias subtraction, flat fielding, wavelength calibration,
and sky subtraction, the spectra were re-aligned to correct for
dithering and stacked. Since the shift values were set to an integer
number of pixels, no re--sampling was applied, in order not to
affect the noise statistics. The stacking was performed using a
simple median filter to reject cosmic rays.

\begin{figure*}[t]
\includegraphics[width=12cm,angle=90]{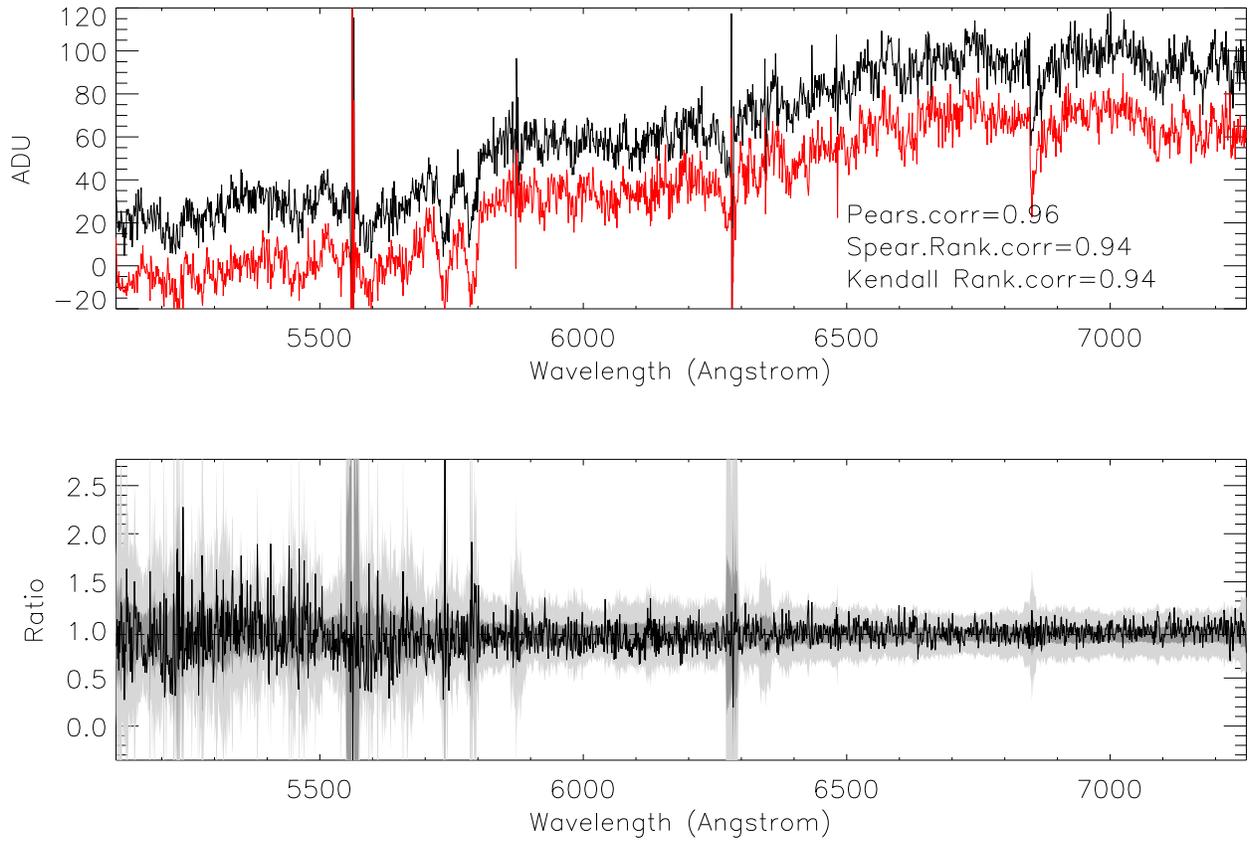}
 \caption{Upper panel: the spectra of the two components of CSL-1.
 Lower panel The ratio between the two spectra in the upper panel.
 The $1$ and $3\sigma$ limits are shown
 as dark and light grey shaded regions.}\label{spectra}
\end{figure*}

The spectra of each component was extracted using a 5 pixel strip (1
pix=0.21") centered on the emission peak with the purpose of maximizing the
S/N ratio. In a similar way, the background counts were extracted in
two stripes located 40 pixels from each component so to
measure the local background while minimizing the contribution from the
source. The error on the spectral counts was calculated with the
following expression:
$$\sigma(ADU)=\sqrt{\sigma_{bkg}^2 + \frac{N(ADU)}{n_{exp}
\times g}}$$
where: $N(ADU)$ are the source counts measured along the
spectrum, $n_{exp}$ are the number of median averaged exposures, $g$
is the instrumental gain, and $\sigma_{bkg}$ represents the background
r.m.s. measured over $5\times 20$ pixels centered at each wavelength.
By folding the images we estimate an average cross-contamination of $7\%\pm 1\%$.
The resulting spectra and their ratio are shown in
Figure~\ref{spectra}. No flux calibration was applied to the data.
It has to be stressed that the narrow spikes visible in the Figure
are residuals left after the removal of bright sky lines.

The spectra of the two components turned out to be identical at visual 
inspection. In fact Pearson's, Spearman, and Kendall correlation tests indicate
correlation coefficients of $0.96$, $0.94$ and $0.94$ respectively, with a significance
of $>99.9\%$ in all cases.
The degree of similarity was further quantified by running a $\chi^2$ test
both on the whole wavelength range\footnote{The regions affected by sky lines subtraction residuals 
were excluded from the test.} and on the most prominent spectral absorption features, namely 
Ca II ($H$, $K$), H$\beta$, $H\delta$, $H\gamma$ lines and $G$ band. The
test yields $\chi^2_\nu=1.03$ implying that the two spectra are 
consistent within $<2\sigma$ (80\%); we also compared the distribution of the 
observed differences to the frequencies expected in the case of pure gaussian noise, finding that the two 
are consistent at the 95\% level and that there is no deterministic part in the residuals. 
An even better agreement ($1\sigma$) is found for the individual absorption features.
To further check if the observed consistency can be due to the known similarity of
early-type galaxies we repeated the test on a sample of spectra extracted from the 
SDSS\footnote{Sloan Digital Sky Survey, data release 4: http://www.sdss.org} {\it Luminous Red Galaxies}, chosen to have a redshift difference and S/N comparable with the CSL1 data. We performed 2000 comparisons obtaining that $<2\%$ of the examined SDSS spectra are as consistent as the spectra of the two CSL1 components. 

A cross correlation test based on the spectral 
lines mentioned above yelds a velocity difference between the two components of 
$\Delta v=14\pm 30$ Km s$^{-1}$. If however we exclude the $H\beta$ line which is 
affected by residual instrumental effects this figure reduces to $0 \pm 20$ Km s$^{-1}$.

\section{Conclusion}

The similarity of the spectra of the two components of CSL-1 and the
zero velocity shift between them strongly support the
interpretation of CSL-1 in terms of gravitational lensing.
The data obtained so far do not allow to completely rule
out the possibility of a chance alignment of two giant ellipticals
but the new data presented in this paper make such an
alignment very unlikely. 
In the case of chance alignment, in fact, the two images of CSL-1
would correspond to two giant ellipticals with identical shapes and
spectra, placed at the same redshift. 
The probability of finding two ellipticals of $M_R=-22.3$ within 2" (20 kpc) and with a radial distance 
$<1$ Mpc ($2\sigma$ upper limit) is $P\sim 1.5\times 10^{-15}$,
accounting for clustering effects \citep[e.g.][]{zeh}\footnote{$P=\int_{V_1,V_2} N^2_{gal}[1+\xi(r)]dV_1 dV_2$ were 
$N_{gal}$ is the space density of elliptical galaxies, $\xi(r)$ is the galaxy correlation function, 
$V_1$ is the volume enclosing the two galaxies and $V_2$ is the volume of the survey.}. 
Integrating over the volume sampled by the OAC-Deep Field for a galaxy of the same magnitude as CSL-1 we calculate that we expect to find $\sim 9\times 10^{-4}$ pairs in the whole survey. Including the spectral similarity we obtain an upper limit of $P<2\times 10^{-5}$.
As it was already mentioned in
Paper I, this could still be explained if CSL-1 belonged to a rich cluster with two central
dominant galaxies, but this is not the case. Careful inspection of the
CSL-1 field shows in fact that it is a rather isolated object with
no other nearby galaxies of comparable brightness; furthermore the 
velocity difference measured from the two spectra is much smaller 
that the one expected in a rich cluster ($\Delta v\sim 300$ Km s$^{-1}$).

In the gravitational lensing scenario, as already stated in Paper I, the observed phenomenology
cannot be understood in terms of lensing by compact clumps of matter such as,
for instance, a Singular Isothermal Sphere model or any other model
listed in the C.R. Keeton's Lens Modeling Software \citep{kee}. 
The only possible type of lens which can produce a morphology similar to
that observed in CSL-1 seems to be a cosmic string \citep{csl1}. 

Lensing by a cosmic string seems capable to
explain all the observational evidences gathered so far and deserves
further investigation. Cosmic strings were predicted by 
\cite{kib76} and their role in cosmology has been
extensively discussed by \cite{zel} and 
\cite{vil}. Recent work \citep{kibble, pol, dav05} has also shown
their relevance for both fundamental physics and cosmology. In
particular it has become apparent that the detection of a cosmic
string would lead to a direct  measure of the energy scale of
symmetry breaking in GUT theories. If we assume that CSL-1 is
produced by a cosmic string, its measured properties would imply a
linear density of the string of order of $G\mu \approx 4\cdot
10^{-7} $ and a corresponding energy scale of GUT of $\sim 10^{15}$
GeV \citep{kib76, pdg}.

Hopefully the question on the nature of CSL1 will soon be answered by
our HST observations approved in Cycle 14 to carry out the 
test proposed by the authors in Paper I and which will allow to verify
the cosmic string hypothesis on firmer grounds.

\section*{Acknowledgments}

M.V.Sazhin acknowledges the INAF-Capodimonte Astronomical Observatory for
hospitality and financial support. O. Khovanskaya acknowledges the
Department of Physics of the University Federico II in Naples for
financial support. The authors wish to thank C. Cezarsky, Director General
of the European Southern Observatory for allocating Director's
Discretionary Time to this project. The work was also supported by
the Russian Fund of Fundamental Investigations No. 04-02-17288. We thank the anonymous 
referees for the helpful suggestions.

\clearpage

\clearpage

\clearpage

\end{document}